\documentclass[mathleft
]{an}
\usepackage{graphicx}
\usepackage{times}
\begin{document}

\Pagespan{1}{3}
\Yearpublication{2009}%
\Yearsubmission{2008}%
\Month{}%
\Volume{}%
\Issue{}%

\title{RAVE -- First Results and 2$^{nd}$ Data Release}

\author{A. Siebert\inst{1}\fnmsep\thanks{Corresponding author:
           \email{siebert@astro.u-strasbg.fr}}
        \and the RAVE collaboration
}

\titlerunning{RAVE 2$^{nd}$ data release and first results}

\authorrunning{A. Siebert \& the RAVE collaboration}

\institute{
Universit\'e de Strasbourg, Observatoire Astronomique, 
F-67000 Strasbourg, France
}

\received{}
\accepted{}
\publonline{later}

\keywords{}

\abstract{ 
RAVE, the RAdial  Velocity Experiment, is an ambitious  program to conduct a
survey to measure the  radial velocities, metallicities and abundance ratios
for  up to  a million  stars using  the 1.2-m  UK Schmidt  Telescope  of the
Anglo-Australian Observatory (AAO), over the  period 2003 - 2010. The survey
represents a  giant leap forward in  our understanding of our  own Milky Way
galaxy, providing  a vast stellar  kinematic database larger than  any other
survey proposed  for this coming  decade.  The main  data product will  be a
southern  hemisphere survey  of about  a million  stars.  This  survey would
comprise  0.7 million  thin disk  main  sequence stars,  250\,000 thick  disk
stars,  100\,000 bulge  and  halo stars,  and  a further  50\,000 giant  stars
including some out to  10 kpc from the Sun. RAVE will  offer the first truly
representative  inventory  of  stellar   radial  velocities  for  all  major
components of the  Galaxy.  Here we present the  first scientific results of
this survey as  well as its second data release which  doubles the number of
previously released radial velocities.  For the first time, the release also
provides atmospheric  parameters for a  large fraction of the  2$^{nd}$ year
data making  it an unprecedented  tool to study  the formation of  the Milky
Way.}

\maketitle


\section{Introduction}

It is now widely accepted that the Milky Way galaxy is a suitable laboratory
to study the formation and evolution of galaxies.  Despite the fact that the
Galaxy  is one unique  system, understanding  its formation  holds important
keys to understand  the broader context of disc  galaxy formation. Thanks to
the past and ongoing large surveys  such as Hipparcos, SDSS, 2MASS or DENIS,
we have access  to data which contains a wealth  of information which enable
us to refine our knowledge of Galaxy formation.  However, with the exception
of  the  SDSS survey,  the  full  description of  the  6D  phase space,  the
combination of the position and velocity spaces, is not available due to the
missing radial velocity and/or distance.

With the advent of multi-fiber  spectroscopy, combined to the large field of
view of  Schmidt telescopes,  it is  now possible to  acquire spectra  for a
large sample of  stars representative of the populations of  the Galaxy in a
reasonable amount of time. Spectroscopy  enables us to measure the generally
missing  radial  velocity, which  allows  us to  study  the  details of  the
dynamics of a system.  Spectroscopy also permits to measure the abundance of
chemical elements in  a stars atmosphere which holds  important clues on the
chemical  evolution of  its parent  system.  Measuring  this  missing radial
velocity and the chemical  abundances to complement the existing astrometric
and photometric catalogues is the main purpose of the RAVE project, with the
driving goal to understand how the Milky Way formed.

RAVE uses the 6dF, the multi--fiber spectroscopic facility at the UK Schmidt
telescope   of   the  Anglo--Australian   Observatory   in  Siding   Spring,
Australia. The  6dF enables us  to collect up  to 150 spectra in  one single
pointing, with a  resolution of 7\,500 in the  Calcium triplet region around
8\,600\AA. This medium resolution  allows the measurement of accurate radial
velocities  ($\sim$2~km/s)  as well  as  atmospheric parameters  ($T_{eff}$,
$\log g$, [M/H]) and chemical abundances.  In Section 2 we present the first
scientific outcomes obtained  using the RAVE data, while  Section 3 presents
the second data release of the  RAVE project (DR2) and discusses the current
status and prospects of the project.

\section{First Results from the Survey}

\noindent{\bf  Escape Velocity}

\noindent  Smith et  al.  (2007)  revisited the  local escape  speed  of our
Galaxy combining a sample of high  velocity stars (HVS) detected by the RAVE
survey to previously known HVS.   Using a maximum likelihood technique, they
find  \newline \mbox{$498<v_{esc}<608$~km/s}  at the  90\%  confidence level
with a median likelihood of 544~km/s.  This result demonstrates the presence
of a dark halo in the Milky Way.\\

\noindent{\bf  Streams in  the Solar  Neighborhood:} 

\noindent Seabroke  et al. (2008) searched  the CORAVEL and  RAVE survey for
infalling stellar streams in the solar vicinity.  Using a Kuiper statistics,
they demonstrated  that the solar  neighborhood is empty of  vertical stream
containing hundreds of stars.  Therefore, it confirms the latest simulations
predicting the  absence of  the Sgr stream  near the  Sun as well  as argues
against  the  Virgo  overdensity  crossing  the  disc  close  to  the  solar
neighborhood.\\

\noindent{\bf Vertical  Structure of the Galactic  Disc:}

\noindent Using  G and  K type stars  towards the Galactic  poles, combining
RAVE to  photometric and astrometric  catalogues, Veltz et al.   (2008) were
able to identify a kinematic discontinuity between the thin and thick discs.
The  existence  of  such a  discontinuity  is  a  strong constraint  on  the
formation scenario  of the thick disc, arguing  against continuous processes
such as the scattering by  spiral arms or molecular arms.  Violent processes
such  as  the  accreted  component  model or  violent  heating  are  instead
preferred.\\

\noindent{\bf Diffuse  Interstellar Bands:}

\noindent  Munari et  al. (2008)  used spectra  of hot  stars from  the RAVE
survey to investigate the properties  of 5 diffuse interstellar bands (DIB).
Their findings indicate that the DIB at 8620.4\AA\,\, is strongly correlated
to reddening and follows  the relation $E(B-V)=2.72(\pm0.03) EW$(\AA), where
$EW$ is the equivalent width of the DIB. It makes this DIB a suitable tracer
of reddening in  the spectra of hot stars. On the  other hand, the existence
of the DIB at 8648\AA\,\, is confirmed but its intensity or equivalent width
does not appear to correlate with reddening.\\

\noindent{\bf Tilt  of the Velocity  Ellipsoid:}

\noindent  Using RAVE  red clump  giants  towards the  South Galactic  pole,
Siebert et al.  (2008) measured the inclination of the velocity ellipsoid at
1~kpc below the Galactic plane. The value of the tilt, $7.3\pm 1.8^{\circ}$,
is consistent with either a short scale length for the disc ($R_d\sim2$~kpc)
if the halo is oblate, or a  long scale length ($R_d\sim 3$~kpc) if the halo
is prolate.   Combined to  independent measurements of  the minor--to--major
axis ratio of the halo which prefers an almost round halo, a scale length of
the disc in the range $[2.5-2.7]$~kpc is preferred.

\section{DR2, Current Status and Ongoing efforts}

{\bf Second Data Release}

\noindent  In  July  2008,  the  RAVE collaboration  released  its  2$^{nd}$
catalogue  (Zwitter  et  al.    2008).   This  second  data  release  covers
$\sim$7\,200 $\deg^2$  in the southern hemisphere and  at least $25^{\circ}$
away from the  Galactic plane.  As for the first  data release (Steinmetz et
al.  2006), the  magnitude range is $9 \leq I_{IC}  \leq 12$, where $I_{IC}$
is the input  catalogue $I$ magnitude, derived from  Tycho-2 $B_T$ and $V_T$
magnitudes for  the bright sample of the  survey ($I_{IC} < 11$)  and is the
SuperCosmos photographic $I$ for the faint sample ($11< I_{IC} < 12$).

DR2  contains 51\,829  radial velocity  measurements for  49\,327 individual
stars, with a typical radial velocity error for the new data between 1.3 and
1.7~km/s.  This catalogue doubles the number of radial velocities previously
published by  the collaboration in  the first data release.   RAVE continues
its  calibration  campaign and  the  comparison  to  external data  shows  a
standard deviation of 1.3~km/s for  the radial velocities, which is twice as
good as for the first data release.

In  addition  to  radial  velocities,  the  catalogue  contains  atmospheric
parameters,  $\log g$,  [M/H]  and  $T_{eff}$, for  the  first time.   These
parameters could  been measured on 22\,407 spectra  corresponding to 21\,121
individual stars. Conservative error  estimates for these parameters for the
average signal--to--noise ratio of the survey (S/N$\sim$40) is 400~K for the
effective temperature, 0.5  dex for the logarithm of  the gravity ($\log g$)
and 0.2 dex for the metallicity. These errors depend strongly on the S/N and
stars  at  the  extreme  of  the  S/N  range  have  errors  $\geq  2$  times
better/worse.   The  calibration  of  the  RAVE  atmospheric  parameters  is
obtained using a specific validation campaign where high--resolution spectra
of standard stars have been taken with various instruments as well as medium
resolution 6dF spectra.

The       catalogue       is       available       from       the       RAVE
website\footnote{http://www.rave-survey.org} or at  the CDS using the VizieR
database.\\

\noindent {\bf Pilot Survey}

\noindent The  pilot survey is now  completed with a  data release scheduled
for mid-2009. It contains about 85\,000 radial velocity measurements as well
as measurements  of stellar atmospheric  parameters for a large  fraction of
the sample. The  release of the pilot survey will mark  a major step forward
in the  RAVE project, and  subsequent data releases  will be based on  a new
input catalogue drawn from the DENIS survey.\\

\noindent {\bf Current Status}

\begin{figure*}
\centering
\includegraphics[width=20cm,angle=270]{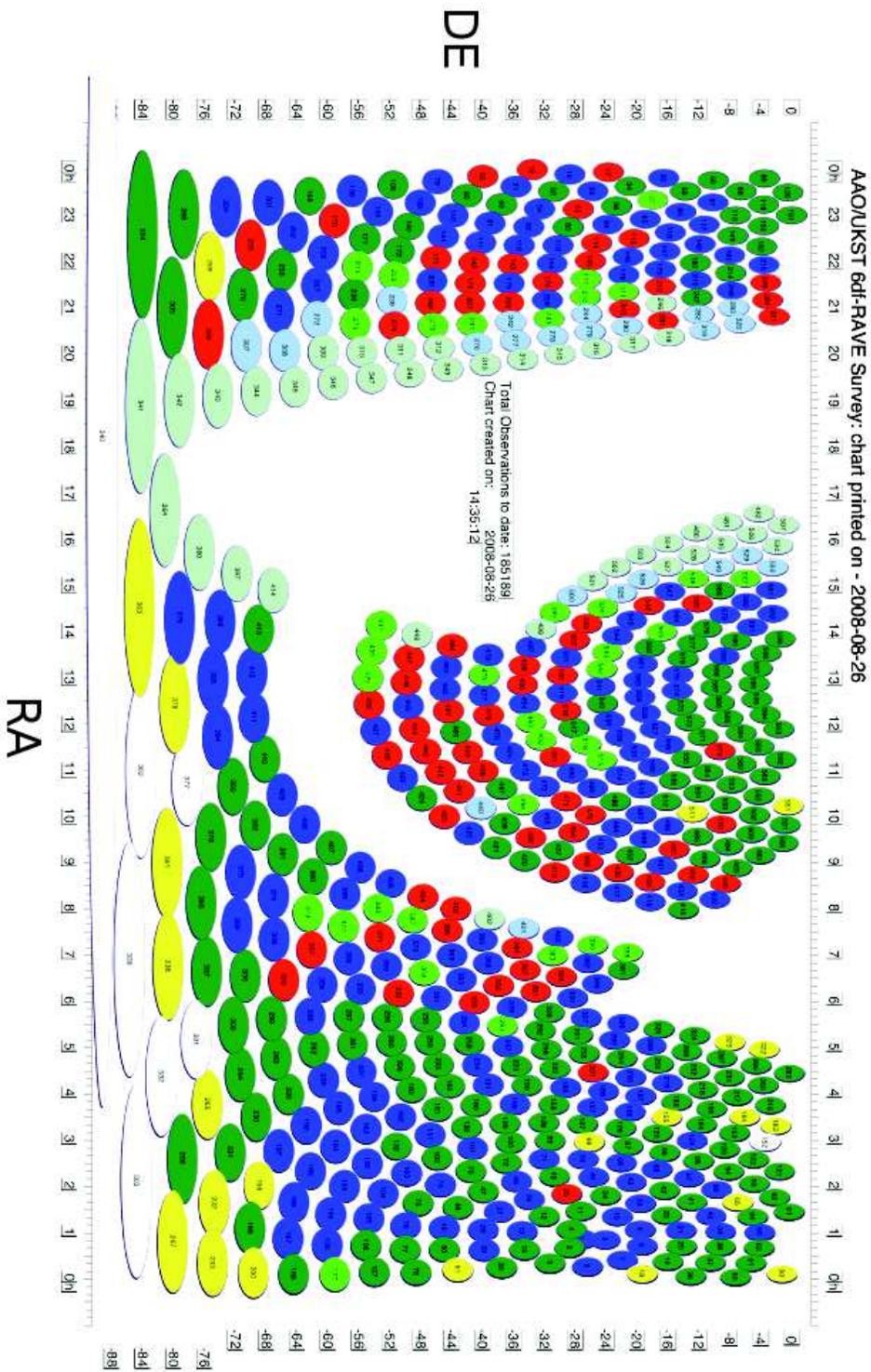}
\caption{Observing chart  of the RAVE  main survey as  of 2008 August  26 in
  equatorial  coordinates.   The figure  is  color-codded  to represent  one
  (yellow), two (green),  three (blue), four (red), five  (light green), six
  (pale blue) or more (grey) pointings of a RAVE field.  The total number of
  spectra  collected for  the main  survey is  185\,189 for  184\,950 unique
  stars. (Figure kindly provided by Fred Watson, AAO)}
\label{f:main}
\end{figure*}

\noindent RAVE currently operates on its full potential, using 25 nights per
lunation. The  project entered its main  survey pha\-se in  March 2006. This
main  survey  relies  on a  new  input  catalogue  based on  DENIS  $I$-band
magnitudes. So far, RAVE has collected approximately 185\,000 spectra in its
main phase,  which adds to  a total of  301\,000 spectra for  270\,000 stars
when combining the  two phases of the project. The  observation chart of the
main phase  as of  2008 August 26  is presented in  Fig.~\ref{f:main}.  This
figure  presents a  colored  representation  of the  RAVE  pointings in  the
equatorial  frame. Each  circle is  a RAVE  5.7''in diameter  field  (not to
scale) and the  color-coding follows the number of visit,  up to seven times
(see  caption for  more details).   Seasonal  variation of  the density  can
easily be seen, the region between  15 and 21 hours in right ascension being
better covered than other regions.\\

\noindent {\bf Ongoing Efforts}

\noindent RAVE primary goal is to measure accurate radial velocities for the
stars  in the Southern  hemisphere.  The  quality of  the RAVE  spectra also
enables us to measure atmospheric parameters ($\log g$, [M/H] and $T_{eff}$)
which  are  used  to  select   subsample  of  the  catalogue  with  peculiar
properties.   For   example,  the  combination   of  colors  and   $\log  g$
measurements from  RAVE permits  an accurate selection  of red  clump giants
(Veltz et al.   2008, Siebert et al. 2008). For  this population, the narrow
luminosity function enables us to obtain distances accurate to 20\% from the
apparent  magnitude  alone.  However,  the  measurement  of the  atmospheric
parameters is non-trivial at RAVE resolution and the transformation from the
measured parameters to the true  parameters relies on calibration data.  The
RAVE   collaboration    constantly   acquires   data    for   standard   and
pseudo-standards stars,  using both  the 6dF instrument  and high-resolution
spectra from other instruments (at  ESO, Apache Point, AAO, Asiago and Haute
Provence), with the aim to refine our calibration.\\

Despite the medium  resolution of the RAVE spectra,  chemical abundances for
about 15  elements can be measured  with a good accuracy  ($\sim$0.2 dex) in
the high signal  to noise spectra. The RAVE  collaboration puts a particular
effort on measuring and validating  these abundances. For this purpose, RAVE
also acquires  spectra for nearby  stars for which accurate  abundances have
been measured  using high-resolution  spectroscopy. So far,  abundances have
been  computed  for  more  than  20\,000 RAVE  targets,  and  the  resulting
measurements  will  be  published   in  a  companion  catalogue  (Boeche  et
al. in preparation).\\

The  knowledge of  the 6D  phase space  requires that  we can  transform the
proper motions into  space velocities. This step relies  on the knowledge of
distances for individual stars,  which can be obtained combining atmospheric
parameters and colors  to isochrones. The availability of  distances in the
near future will enable the full potential of the RAVE catalogue, permitting
the study of the detailed shape of the phase space and reveal new details of
the formation of the Solar neighbourhoud.\\

\section{Conclusions}

RAVE released its second catalogue in July 2008, reporting radial velocities
for 51\,829  spectra and 49\,327  different stars, randomly selected  in the
magnitude range of $9<I<12$ and located more than 25$^{\circ}$ away from the
Galactic plane.  This  release doubles the size of  the previously published
catalogue. The typical error on the published radial velocity is between 1.3
and 1.7~km/s.

In addition to radial velocities,  the catalogue contains for the first time
atmospheric    parameters     measurements    for    more     than    20\,000
spectra. Uncertainties for a typical RAVE  star are of 400 K in temperature,
0.5 dex in gravity, and 0.2 dex in metallicity, but the error depends on the
S/N and can vary by a factor of $\geq 2$ for stars at the extreme of the S/N
range.\\

The survey continues  to collect spectra and is  scheduled to continue until
2010.   Currently more  than 300\,000  spectra have  been collected  and are
currently  being processed.   Acquisition of  new calibration  data  is also
underway, enabling  us to refine our atmospheric  parameters measurements as
well  as our radial  velocities, and  represents an  important task  for the
ongoing  effort. Other  ongoing  efforts in  the  collaboration include  the
measurement of chemical  abundances and distances which will  be released as
companion catalogues in the future.\\

The RAVE catalogue can be used to  study the formation of the Milky Way and,
for  example,  the collaboration  succeeded  in  refining significantly  the
measurement of the local escape velocity using the high velocity stars found
in  the RAVE  catalogue. Further  studies carried  out by  the collaboration
include the search for vertical  stream of matter in the Solar neighborhood,
a  new determination of  the vertical  structure of  the Galactic  disc, the
study  of  diffuse  interstellar  bands  in the  Galactic  plane  and  their
correlation  to  the  interstellar  extinction  or the  inclination  of  the
velocity ellipsoid towards the Galactic plane at 1~kpc below the plane.\\

As the survey  progresses and new data become  available, the RAVE catalogue
will enable  us to  refine our view  on the  structure and formation  of the
Milky Way, paving the way for new insights on the formation of galaxies.\\

\acknowledgements Funding for RAVE has been provided by the Anglo-Australian
Observatory,  the Astrophysical Institute  Potsdam, the  Australian Research
Council,  the  German  Research   foundation,  the  National  Institute  for
Astrophysics  at  Padova,  The  Johns Hopkins  University,  the  Netherlands
Research School for Astronomy, the Natural Sciences and Engineering Research
Council of Canada, the Slovenian Research Agency, the Swiss National Science
Foundation, the  National Science Foundation  of the USA  (AST-0508996), the
Netherlands Organisation  for Scientific Research, the  Particle Physics and
Astronomy Research  Council of the UK, Opticon,  Strasbourg Observatory, and
the Universities of Basel, Cambridge, and Groningen. 

\noindent The RAVE Web site is at \mbox{www.rave-survey.org}.


\end{document}